\documentclass[prd,aps,showpacs,preprintnumbers,amssymb]{revtex4}
\usepackage{graphicx}

\usepackage{epsf}

\begin{document}

\title{%
\hfill{\normalsize\vbox{%
\hbox{}
 }}\\
{Probing scalar mesons in semi-leptonic decays of $D_s^+$, $D^+$ and $D^0$}}

\author{Amir H. Fariborz $^{\it \bf a}$~\footnote[1]{Email:
 fariboa@sunyit.edu}}

\author{Renata Jora $^{\it \bf b}$~\footnote[2]{Email:
  rjora@theory.nipne.ro}}

\author{Joseph Schechter $^{\it \bf c}$~\footnote[3]{Email:
 schechte@physics.syr.edu}}

\author{M. Naeem Shahid $^{\it \bf d}$~\footnote[4]{Email:
 mnshahid@physics.syr.edu}}

\affiliation{$^ {\bf \it a}$ Department of Mathematics and Physics,
 State University of New York Institute of Technology, Utica,
 NY 13502, USA}

\affiliation{$^ {\bf \it b}$  National Institute of Physics and Nuclear Engineering PO Box MG-6, Bucharest-Magurele, Romania}

\affiliation{$^ {\bf \it c}$ Department of Physics,
 Syracuse University, Syracuse, NY 13244-1130, USA,}

\affiliation{$^ {\bf \it d}$National University of Science and Technology (NUST)
           H-12 Islamabad, Pakistan}

\date{\today}

\begin{abstract}
With the primary motivation of probing the quark substructure of scalar mesons,  a generalized linear sigma model for the lowest and the next-to-lowest scalar and pseudoscalar mesons is employed to investigate several semi-leptonic decays of $D$ mesons.   The  free parameters of the model (in its leading approximation) have been previously determined from fits to mass spectra and various low-energy parameters.  With these fixed parameters,  the model has already given encouraging predictions for  different low-energy decays and scattering,  as well as for semileptonic decay channels of $D_s^+$ that include a scalar meson in the final state.   In the present work, we apply the same model (in its leading order with the same fixed parameters) to different semi-leptonic decay channels of $D_s^+$, $D^+$ and $D^0$ and thereby further test the model and its predictions for the quark substructure of scalar mesons.    We find that these predictions are in reasonable agreement with experiment.

\end{abstract}


\maketitle

\section{Introduction}

Probing the quark substructure of scalar mesons is proven to be quite non-trivial (see \cite{pdg} - \cite{FJSS11}).  The simple quark-antiquark model, which works well for pseudoscalars and vectors,   does not explain the properties of the scalars below 1 GeV such as their light and inverted mass spectrum.       The MIT bag model of Jaffe provides a theoretical foundation for understanding the properties of lowest-lying scalar mesons within a diquak-antidiquark picture.   While scalars above 1 GeV  are better treated  within the quark model,  they too show some signs of deviations from the quark-antiquark picture \cite{Mec}.     A natural question then arises as to whether various underlying mixings among scalar mesons below and above 1 GeV are in any way responsible for these deviations.     To answer this question it is necessary to investigate the global picture of scalars below and above 1 GeV and study the family relations among them.    This objective was taken up in \cite{global} in which such family relations were studied in some detail within a generalized linear sigma model that includes two nonets of scalar mesons  and two nonets of pseudoscalar mesons (a quark-antiquark nonet and a four-quark nonet).   Prior works also include      \cite{BFMNS01}, \cite{mixing}, \cite{thermo}, \cite{FJS08}, \cite{FJS05}-\cite{global},
     \cite{FJSS11}.

The generalized linear sigma model was also applied to several semi-leptonic decays of $D_s^+(1968)$
measured by the CLEO collaboration \cite{Cleo}.    These included $D_s^+(1968)\rightarrow f_0(980) e^+ v_e$,  as well as $D_s^+(1968)\rightarrow \eta(\eta') e^+ v_e$.   It was shown in \cite{FJSS11} that the model prediction for these semileptonic decays agrees well with the CLEO measurements.    In the present work we study the predictions of the same model for $D_s^+(1968)\rightarrow K^0 e^+ v_e$,   $D^+ \rightarrow  \pi^0 e^+ v_e$,
$D^+ \rightarrow  \eta (\eta') e^+ v_e$, $D^+ \rightarrow  {\bar K}^0 e^+ v_e$,    $D^0 \rightarrow  \pi^- e^+ v_e$ and $D^0 \rightarrow  K^- e^+ v_e$.

In section II we give a brief review of the hadronic ``weak currents" which are needed for the calculation. These are mathematically given by the so-called Noether currents of the sigma model Lagrangian being employed.
We work in the approximation where renormalization of these currents from the symmetry limit are neglected.
This means that there are no arbitrary parameters available to us.
In section III we give a detailed description of the calculation of the partial decay widths from the currents
discussed in section II. For this purpose we also use information on the scalar and pseudoscalar
meson masses and mixings obtained in \cite{global}.  A short summary and discussion is given in section IV.

\section{The Hadronic Vector Currents}
The generalized linear sigma model of ref. \cite{global} describes the global picture of scalar and pseudoscalar mesons below and above 1 GeV in terms of two chiral nonets for scalars and two chiral nonets for pseudoscalars.    These are a quark-antiquark nonet $M$ and a four-quark nonet $M'$ that in turn are expressed in terms of the corresponding scalar and pseudoscalar nonets
\begin{eqnarray}
M &=& S + i \phi,  \nonumber \\
M' &=& S' + i \phi'.
\label{mmp}
 \end{eqnarray}
The hadronic Noether vector currents, that are relevant to the semileptonic decays in the present work are (see Appendix A of \cite{su71}),
\begin{eqnarray}
      V_{\mu a}^b&=&i\phi_a^c\stackrel{\leftrightarrow}{\partial_\mu}\phi_c^b +
    i\tilde{S}_a^c\stackrel{\leftrightarrow}{\partial_\mu}\tilde{S}_c^b
    +i(\alpha_a-\alpha_b)\partial_\mu\tilde{S}_a^b,
\nonumber \\
    V_{\mu a}^{\prime b}&=&i\phi_a^{\prime c}\stackrel{\leftrightarrow}{\partial_\mu}\phi_c^{\prime b} +
    i\tilde{S}_a^{\prime c}\stackrel{\leftrightarrow}{\partial_\mu}\tilde{S}_c^{\prime b}
    +i(\beta_a-\beta_b)\partial_\mu\tilde{S}_a^{\prime b},
\label{su3currents}
\end{eqnarray}
wherein,
 \begin{eqnarray}
        S &=& \tilde{S} + \langle S \rangle,\quad\quad   \langle S_a^b \rangle =\alpha_a \delta_a^b, \nonumber \\
 S^\prime &=& \tilde{S}^\prime + \langle S^\prime \rangle,\quad\quad   \langle {S^\prime}_a^b \rangle  =\beta_a \delta_a^b,
    \label{moresu3vevs}
    \end{eqnarray}
where $\alpha$ and $\beta$ being the vacuum expectation values of fields $S$ and $S'$, respectively.
Therefore,   total vector current is:
\begin{equation}
\left[ V_{\mu a}^b\right]^{\rm tot.}=V_{\mu a}^{b}+V^{\prime b}_{\mu a}.
 \label{totalsu3currents}
\end{equation}
In this framework, the physical states become a linear combination (expressed by a rotation matrix) of the unprimed and primed fields.
The rotation matrices relevant to our study in this work are $R_\pi$, $R_K$ and $R_\eta$ defined as follows:
The
  transformation between the physical $\pi^+$ and $\pi^{\prime +}$ fields and the
  original
fields (say $\phi^+$ and $\phi'^+$) is \cite{FJS05}:
\begin{equation}
\left[
\begin{array}{c}  \pi^+(137) \\
                 \pi'^+(1300)
\end{array}
\right]
=
R_\pi^{-1}
\left[
\begin{array}{c}
                        \phi_1^2 \\
                        {\phi'}_1^2
\end{array}
\right]
\label{R_pi}
\end{equation}
where $R_\pi$ is the pion rotation matrix.
Similarly, for the kaon system
\begin{equation}
\left[
\begin{array}{cc}
K^+(496)\\
{K'}^+(1460)
\end{array}
\right]
=
R_K^{-1}
\left[
\begin{array}{cc}
\phi_1^3\\
{\phi'}_1^3
\end{array}
\right]
\label{R_K}
\end{equation}
and for the eta system
\begin{equation}
\left[
\begin{array}{c}
\eta(547)   \\
\eta(958)  \\
\eta(1295)  \\
\eta(1760)\\
\end{array}
\right]
=
R_0^{-1}
\left[
\begin{array}{cc}
\eta_a\\
\eta_b\\
\eta_c\\
\eta_d
\end{array}
\right],
\label{R_0}
\end{equation}
where
\begin{eqnarray}
\eta_a&=&\frac{\phi^1_1+\phi^2_2}{\sqrt{2}}
\hskip .4cm \rightarrow \hskip .4cm
n{\bar n},
\nonumber  \\
\eta_b&=&\phi^3_3
\hskip 1.25cm \rightarrow \hskip .4cm
 s{\bar s},
\nonumber    \\
\eta_c&=&  \frac{\phi'^1_1+\phi'^2_2}{\sqrt{2}}
\hskip .17cm \rightarrow \hskip .4cm
ns{\bar n}{\bar s},
\nonumber   \\
\eta_d&=& \phi'^3_3
\hskip 1.15cm \rightarrow \hskip 0.4cm
nn{\bar n}{\bar n}.
\label{etafourbasis}
\end{eqnarray}
where on the right the schematic quark substructure is given (in which $n$ stands for non-strange up and down quarks). The rotation matrices have been determined in the global fit of ref. \cite{global} and will be used in the present work.

When extending this model to the case where a heavy flavor (such as the charm quark here) is added,  the heavy spin zero mesons need to be  considered as quark-antiquark states.     This is based on the findings of ref. \cite{FJSS11} in which it is shown that the case of three flavors is special in the sense that
it is the only one in which a two quark-two antiquark field has the correct chiral
transformation property to mix (in the chiral limit) with $M$.    Therefore, in the present generalized  linear sigma model the kinetic term would then be written as:
    \begin{equation}
    {\cal L} = -\frac{1}{2} {\rm Tr}^4(\partial_\mu M \partial_\mu M^\dagger)
               -\frac{1}{2} {\rm Tr}^3(\partial_\mu M^\prime \partial_\mu M^{\prime\dagger}),
      \label{hybridlag}
       \end{equation}
where the meaning of the superscript on the trace symbol is
that the first term should be summed over the heavy quark index as
well as the three light indices. This stands in contrast to the second term
which is just summed over the three light quark indices pertaining to the
two quark - two antiquark field $M^\prime$. Since the Noether
currents are sensitive only to these
kinetic terms in the model, the vector currents with
flavor indices 1 through 3 in this model are just the same as in
Eq.(\ref{totalsu3currents}) above. However if either or both flavor
indices take on the value 4 (referring to the heavy flavor) the current
will only have contributions from the field $M$, i.e.
\begin{equation}
 \left[ V_{\mu 4}^a \right] ^{\rm tot.} = V_{\mu 4}^a=i\phi_4^c\stackrel{\leftrightarrow}{\partial_\mu}\phi_c^a +
    i S_4^c\stackrel{\leftrightarrow}{\partial_\mu}S_c^a,
\label{heavycurrents}
\end{equation}
where the unspecified
indices can run from 1 to 4.     The current is given in terms of the ``bare'' fields which are related to the
physical fields through the non-derivative terms (``potential'') terms of the effective Lagrangian.    The connections between the ``bare'' and physical fields are given by the appropriate rotation matrices discussed above.

\section{Semileptonic $D$ decays}

\subsection{$D_s^+ \rightarrow K^0 e^+ \nu_e$}
\begin{figure}[htbp]
\centering
\rotatebox{0}
{\includegraphics[width=7cm,clip=true]{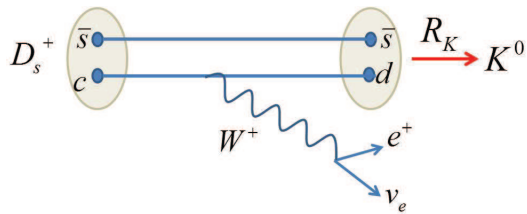}}
\caption[]{
Schematic diagram for semileptonic decay $D_s^+ \rightarrow K^0 e^+ \nu_e$.    The rotation matrix $R_K$ is computed in ref. \cite{global} and projects the produced $d{\bar s}$ pair onto the wave function of $K^0$.
}
\label{Ds_K0ev_FD}
\end{figure}

\begin{figure}[htbp]
\centering
\rotatebox{0}
{\includegraphics[width=7cm,clip=true]{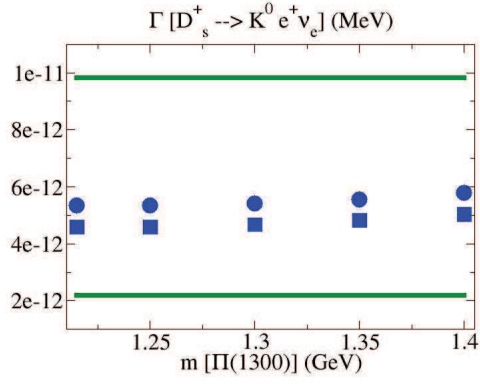}}
\caption{Prediction of the $MM'$ model for the  partial decay width of $D_s^+ \rightarrow K^0 e^+ \nu_e$ versus $m[\pi(1300)]$ for values of $A_3/A_1$ equal to 20 (circles) and 30 (squares).   The horizontal lines show the experimental range \cite{pdg}. }
\label{Ds_K0ev_R}
\end{figure}

The schematic diagram for this decay is shown in  Fig. \ref{Ds_K0ev_FD}.   The decay proceeds via the vector current described in Eq.  (\ref{heavycurrents})
\begin{equation}
V_{\mu 4}^2=i\phi_4^3\stackrel{\leftrightarrow}{\partial_\mu}\phi_3^2 + \cdots =
i (R_K)_{11} D_s^+ \stackrel{\leftrightarrow}{\partial_\mu} {\bar K}^0 + \cdots
\end{equation}
where $R_K$ is the rotation matrix for the isodoublet pseudoscalars defined in (\ref{R_K}) and computed in \cite{global}.   This rotation matrix projects the $d {\bar s}$ pair onto the wave function of the $K^0$.   Using the model parameters found in \cite{global} and following the standard calculation of the decay width (Appendix A),  the prediction of the model is given in Fig. \ref{Ds_K0ev_R} and compared with the experimental bounds for this decay width extracted from PDG \cite{pdg} (the horizontal lines).   Clearly,  the prediction of the model which is plotted versus $m[\pi(1300)]$ and for two values of $A_3/A_1$) falls within the experimental bounds.   This provides further support for the effectiveness of the model and its predictions for the quark substructure of scalar and pseudoscalar mesons below and above 1 GeV \cite{global}.

\subsection{$D^+ \rightarrow \pi^0 e^+ \nu_e$, $D^+ \rightarrow \eta(\eta') e^+ \nu_e$ and $D^+ \rightarrow \eta(\eta') e^+ \nu_e$}

\begin{figure}[htbp]
\centering
\rotatebox{0}
{\includegraphics[width=7cm,clip=true]{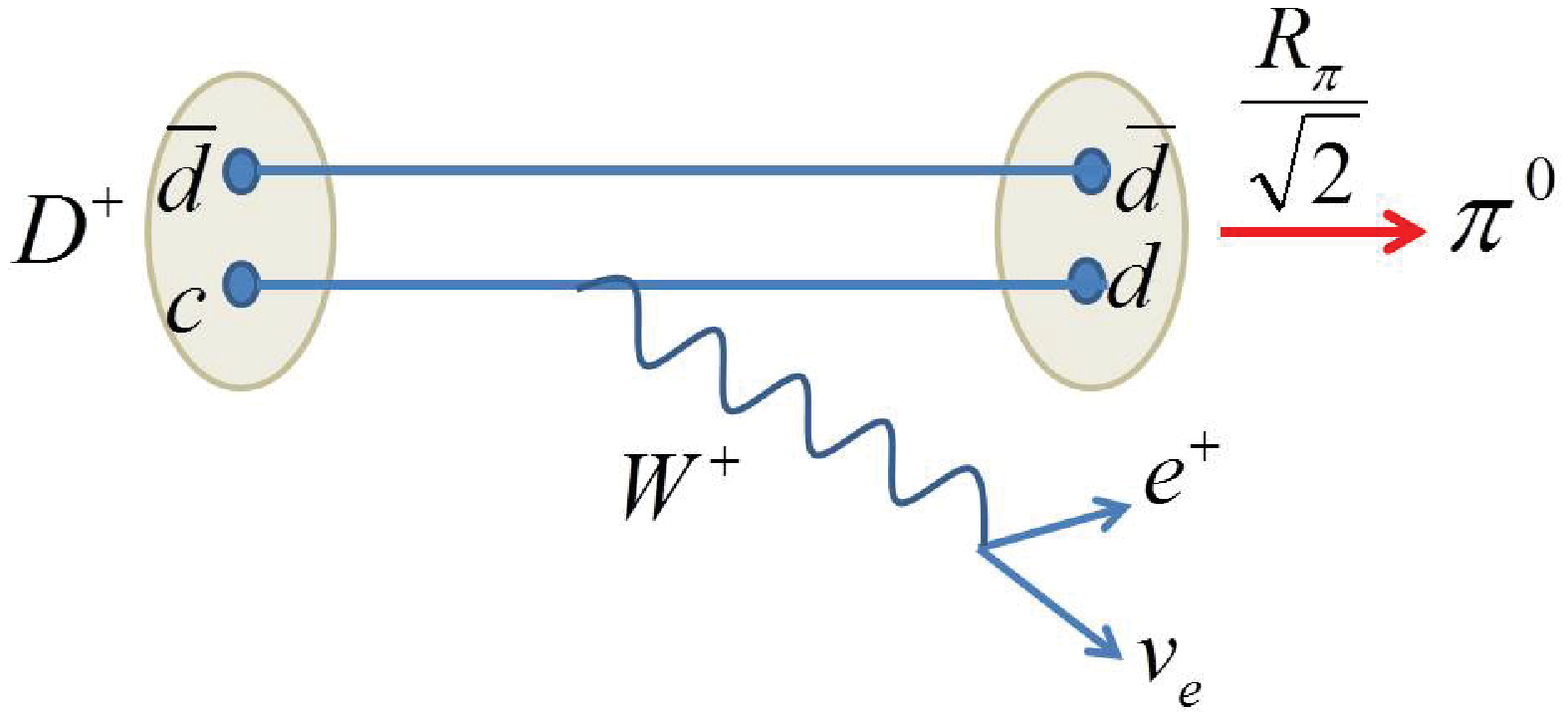}}
\hskip 1.5cm
{\includegraphics[width=7cm,clip=true]{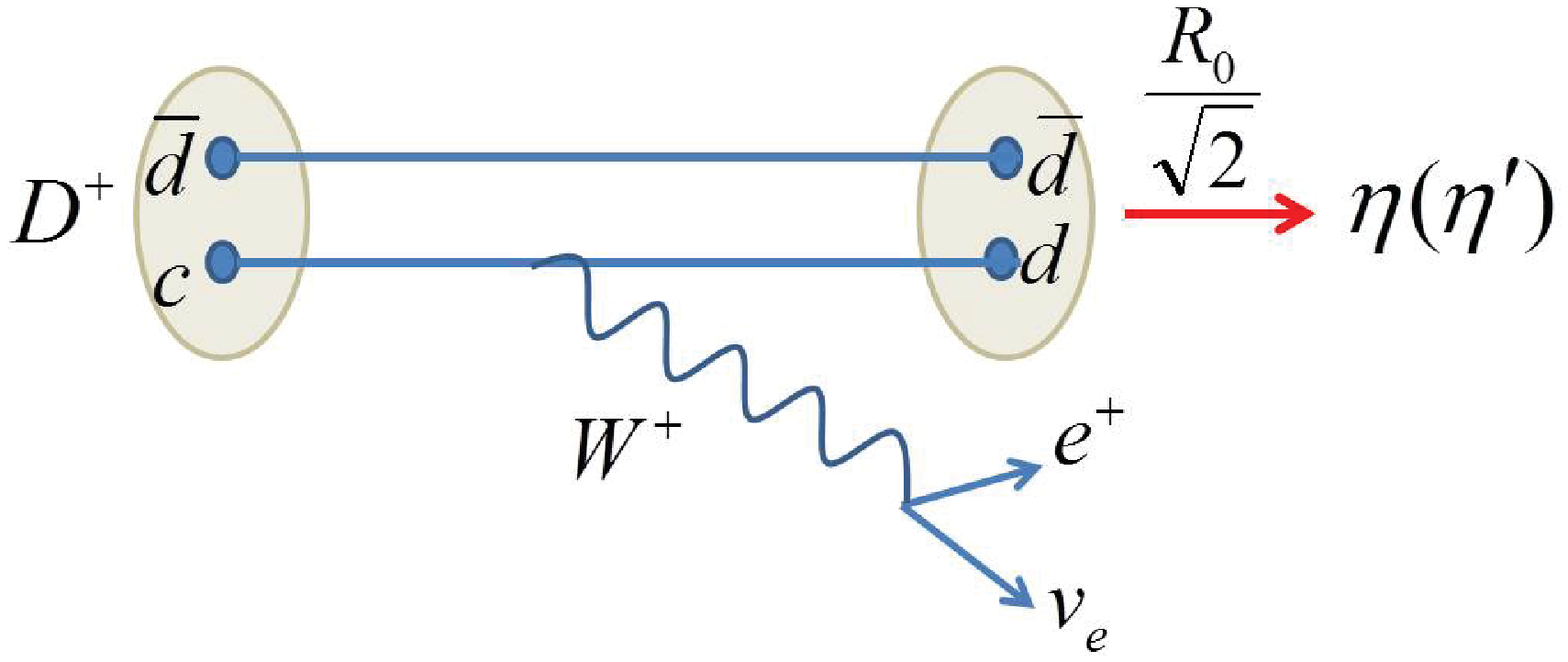}}
\vskip1cm
{\includegraphics[width=7cm,clip=true]{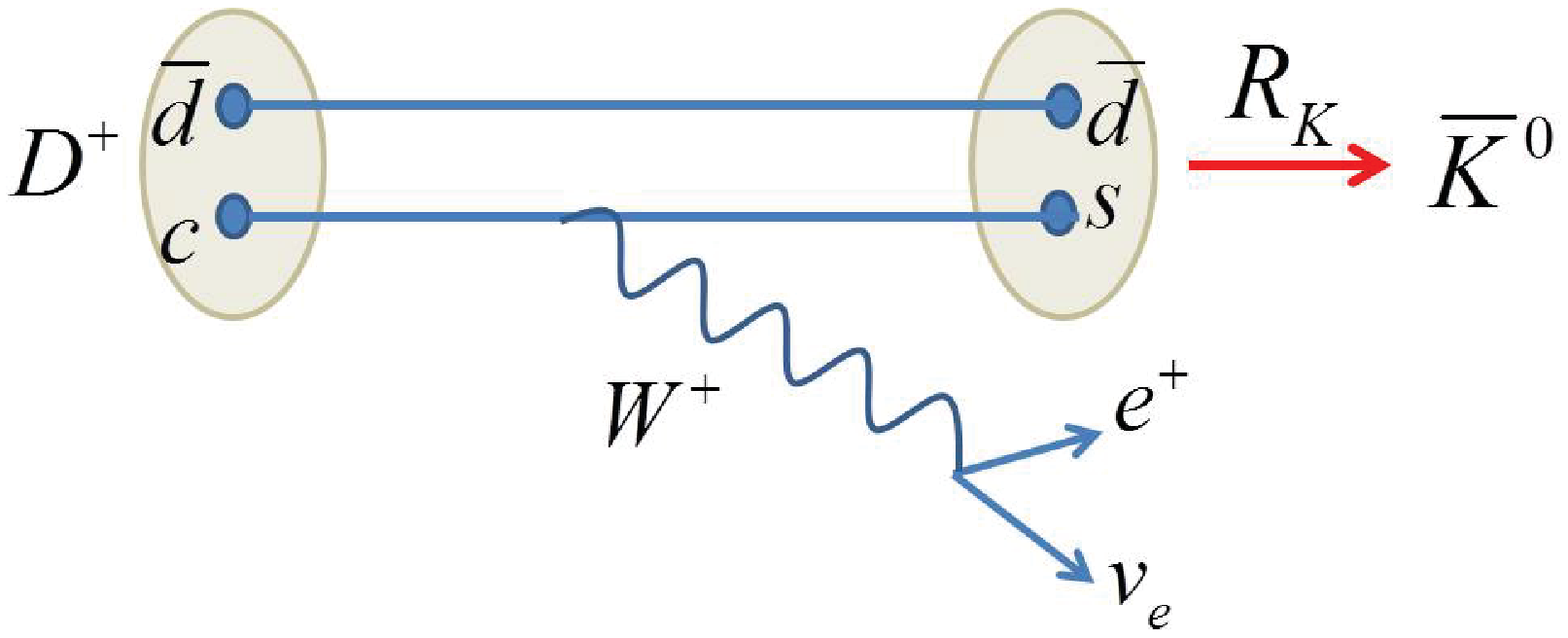}}
\caption[]{
Schematic diagrams for semileptonic decays $D^+ \rightarrow \pi^0 e^+ \nu_e$ (top left);  $D^+ \rightarrow \eta (\eta') e^+ \nu_e$ (top right); and $D^+ \rightarrow {\bar K}^0 e^+ \nu_e$ (bottom).    The rotation matrices $R_0$, $R_\pi$ and $R_K$ are computed in ref. \cite{global} and project the produced $q{\bar q}$ pair onto the wave function of the final state meson.
}
\label{F_D+_FD}
\end{figure}

\begin{figure}[htbp]
\centering
\rotatebox{0}
{\includegraphics[width=7cm,clip=true]{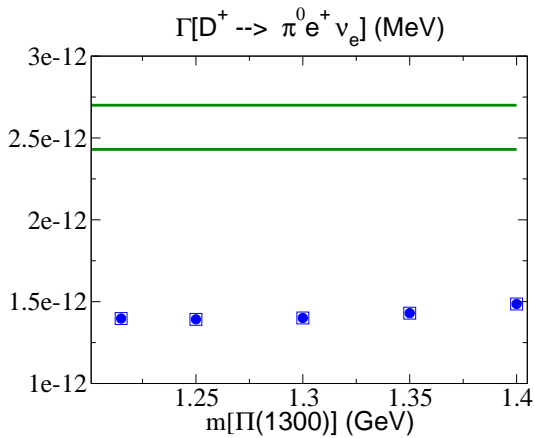}}
\caption{Prediction of the $MM'$ model for the  partial decay width of $D^+ \rightarrow \pi^0 e^+ \nu_e$ versus $m[\pi(1300)]$ for values of $A_3/A_1$ equal to 20 (circles) and 30 (squares).   The horizontal lines show the experimental range \cite{pdg}.}
\label{D+_pi0ev_R}
\end{figure}

\begin{figure}[htbp]
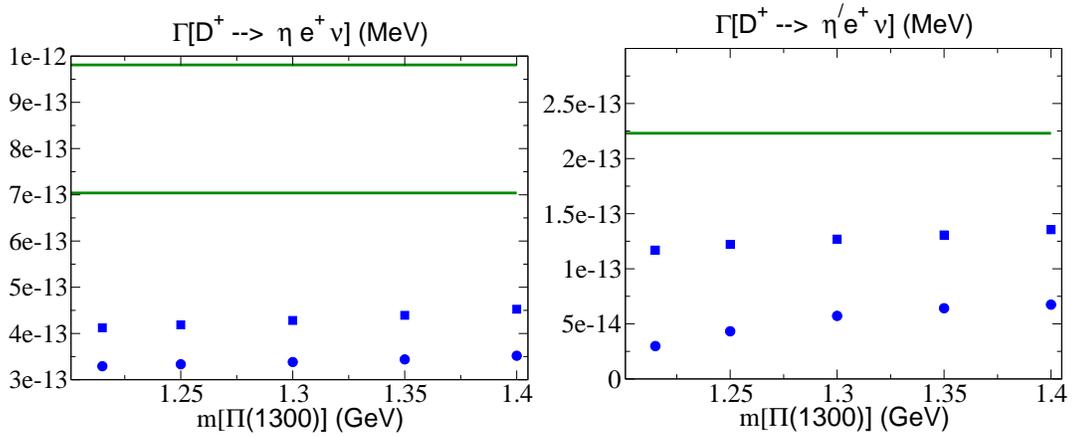

\centering
\rotatebox{0}
{\includegraphics[width=7cm,clip=true]{fig5a.eps}}
{\includegraphics[width=7cm,clip=true]{fig5b.eps}}
\caption{Prediction of the $MM'$ model for the  partial decay width of $D^+ \rightarrow \eta e^+ \nu_e$ (left) and $D^+ \rightarrow \eta' e^+ \nu_e$ (right) versus $m[\pi(1300)]$ for values of $A_3/A_1$ equal to 20 (circles) and 30 (squares).   The horizontal lines show the experimental range \cite{pdg} (for the $\eta'$ channel only upper experimental bound is known).}
\label{D+_etaev_R}
\end{figure}

\begin{figure}[htbp]
\centering
\rotatebox{0}
{\includegraphics[width=7cm,clip=true]{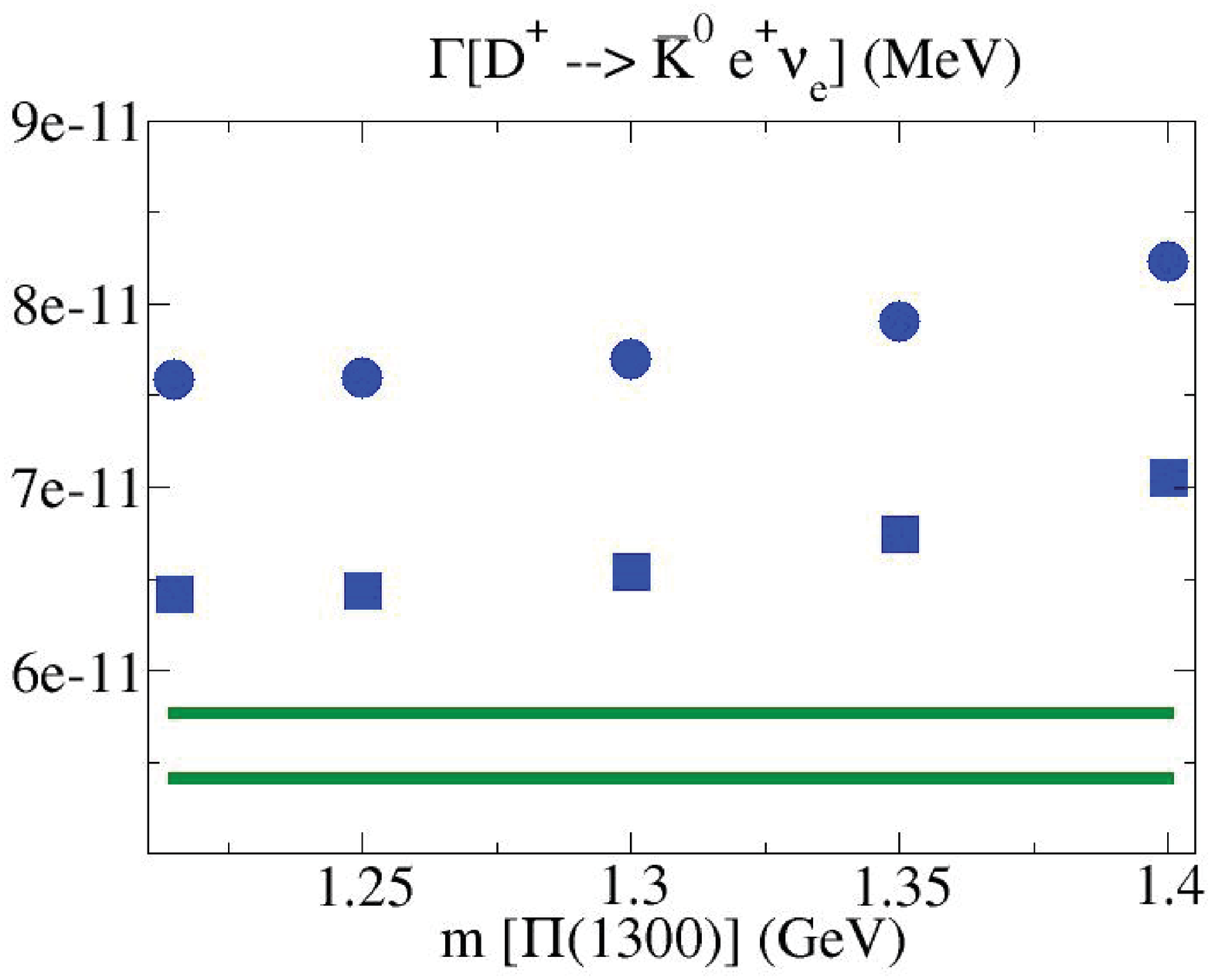}}
\caption{Prediction of the $MM'$ model for the  partial decay width of $D^+ \rightarrow {\bar K}^0 e^+ \nu_e$ versus $m[\pi(1300)]$ for values of $A_3/A_1$ equal to 20 (circles) and 30 (squares).   The horizontal lines show the experimental range \cite{pdg}.}
\label{D+_K0ev_R}
\end{figure}

The schematic diagrams for these  decays are shown in  Fig. \ref{F_D+_FD} and proceed via production of a ${\bar d} d$ pair described by the vector current of Eq.  (\ref{heavycurrents}).   The ${\bar d} d$ pair then gets projected \cite{global} onto the wave function of the final meson by the appropriate rotation matrix $R_K$ (first decay), $R_0$ (second decay) and $R_K$ (last decay).  The relevant vector current for the $\pi^0$ channel is
\begin{equation}
V_{\mu 4}^2=i\phi_4^2\stackrel{\leftrightarrow}{\partial_\mu}\phi_2^2 + \cdots =
-i {1\over\sqrt{2}} \left(R_\pi\right)_{11} D^+  \stackrel{\leftrightarrow}{\partial_\mu} \pi^0 + \cdots
\end{equation}
The extra $\sqrt{2}$ stems from the fact that
\begin{equation}
\pi^0 \propto \left(R_\pi^{-1}\right)_{11} {{u{\bar u} - d {\bar d}}\over \sqrt{2}}
\end{equation}
Therefore the projection of the $d {\bar d}$ onto the $\pi^0$ will include a division by $\sqrt{2}$.     Similarly,   the relevant vector current for the $\eta$ ($\eta'$) channel is
\begin{equation}
V_{\mu 4}^2 = i\phi_4^2\stackrel{\leftrightarrow}{\partial_\mu}\phi_2^2 + \cdots =
-i {1\over\sqrt{2}} \left(R_0\right)_{11(2)} D^+  \stackrel{\leftrightarrow}{\partial_\mu} \eta(\eta') + \cdots
\end{equation}
and for the ${\bar K}^0$ channel
\begin{equation}
V_{\mu 4}^3 = i\phi_4^2\stackrel{\leftrightarrow}{\partial_\mu}\phi_2^3 + \cdots =
i \left(R_K\right)_{11} D^+  \stackrel{\leftrightarrow}{\partial_\mu} K^0 + \cdots
\end{equation}
The prediction of the model for the decay width in the $\pi$ channel is given in Fig. \ref{D+_pi0ev_R} showing an order of magnitude agreement with experiment.    The predictions for the decay width in the $\eta/\eta'$ channels are displayed in Fig. \ref{D+_etaev_R}.   For the $\eta$-channel the agreement is at the level of order of magnitude,  while for the $\eta'$ channel, for which only the upper experimental bound is given in PDG \cite{pdg}, the prediction is within the experimental range.  Similarly,  for the ${\bar K}^0$ channel (Fig. \ref{D+_K0ev_R}) the agreement with experiment is at the level of order of magnitude.

\subsection{$D^0 \rightarrow \pi^- e^+ \nu_e$ and $D^0 \rightarrow K^- e^+ \nu_e$}

\begin{figure}[htbp]
\centering
\rotatebox{0}
{\includegraphics[width=7cm,clip=true]{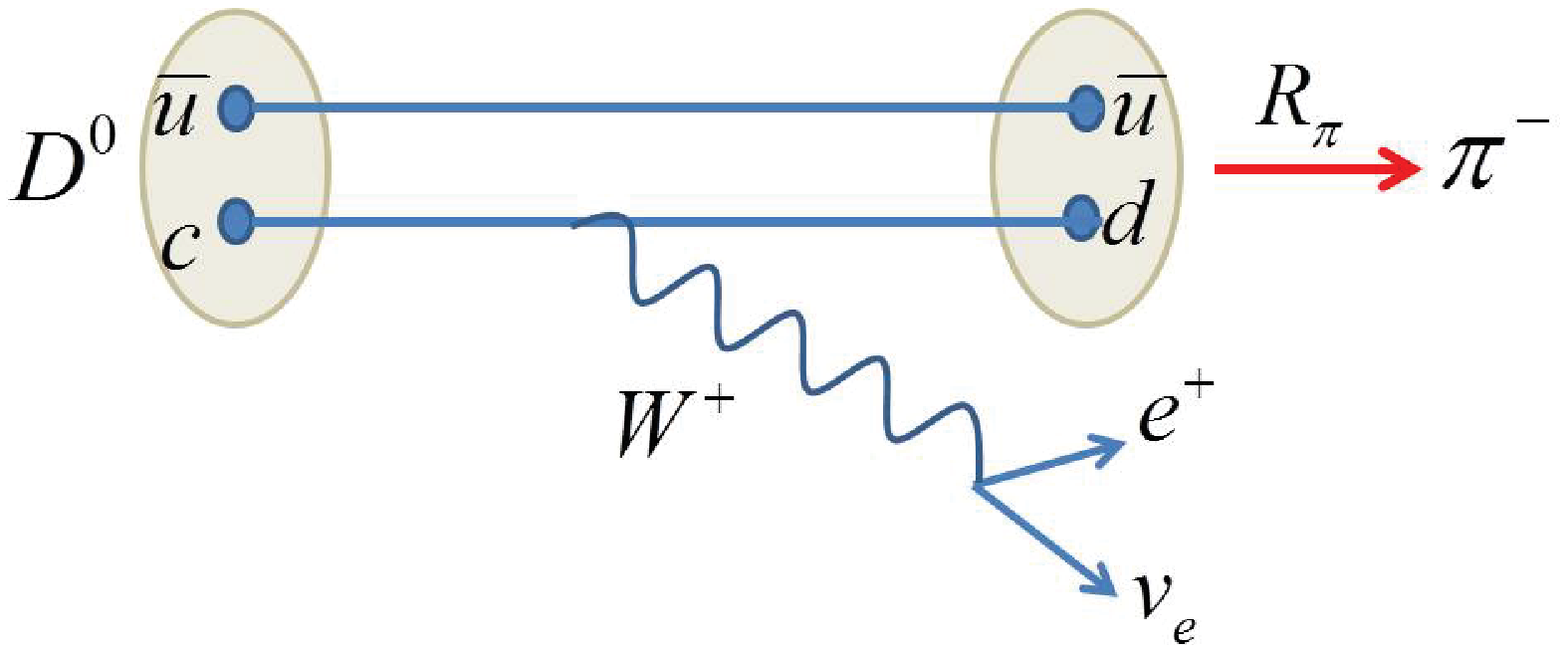}}
\hskip 1.5cm
{\includegraphics[width=7cm,clip=true]{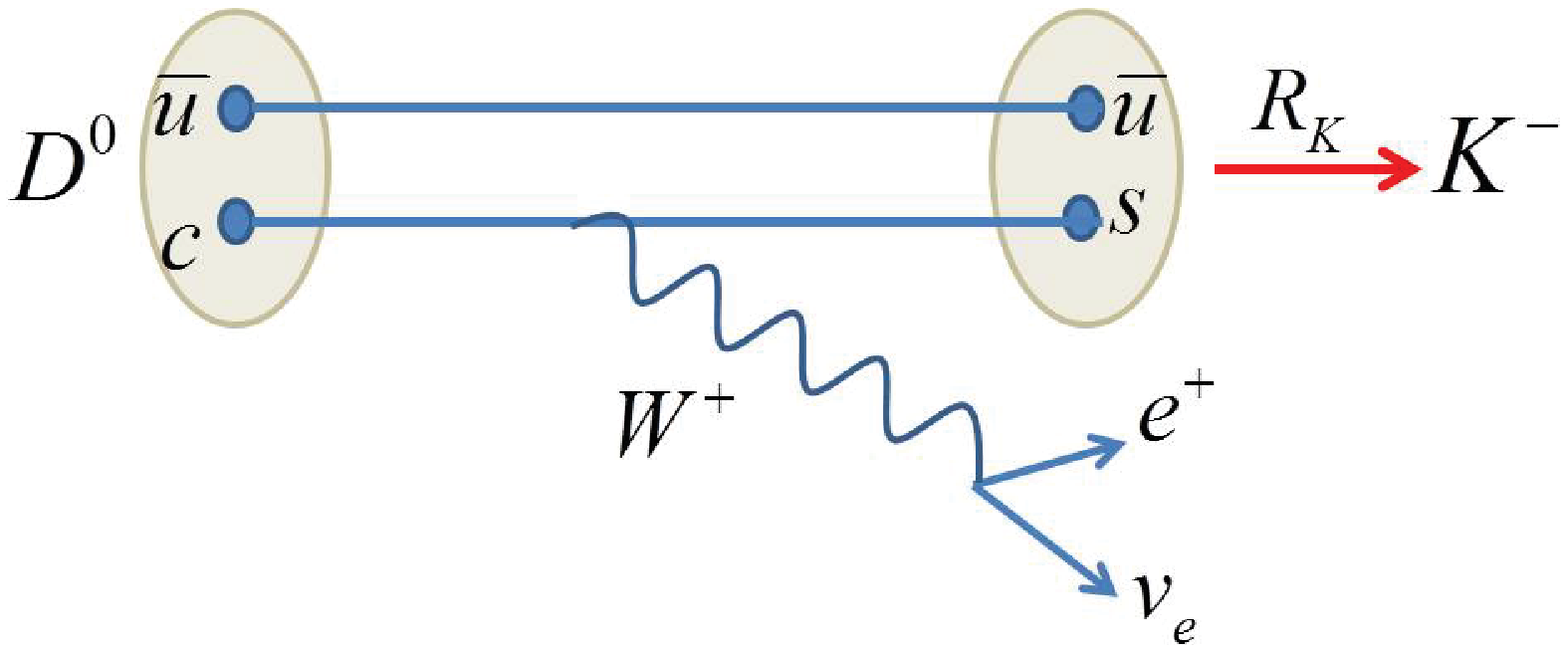}}
\caption[]{
Schematic diagrams for semileptonic decays $D^0 \rightarrow \pi^- e^+ \nu_e$ and $D^0 \rightarrow K^- e^+ \nu_e$.    The rotation matrices $R_\pi$ and $R_K$ are computed in ref. \cite{global} and project the produced $d{\bar u}$ and $s{\bar u}$ pairs onto the wave functions of $\pi^-$ and $K^-$, respectively.
}
\label{D0_piKev_FD}
\end{figure}

\begin{figure}[htbp]
\centering
\rotatebox{0}
{\includegraphics[width=7cm,clip=true]{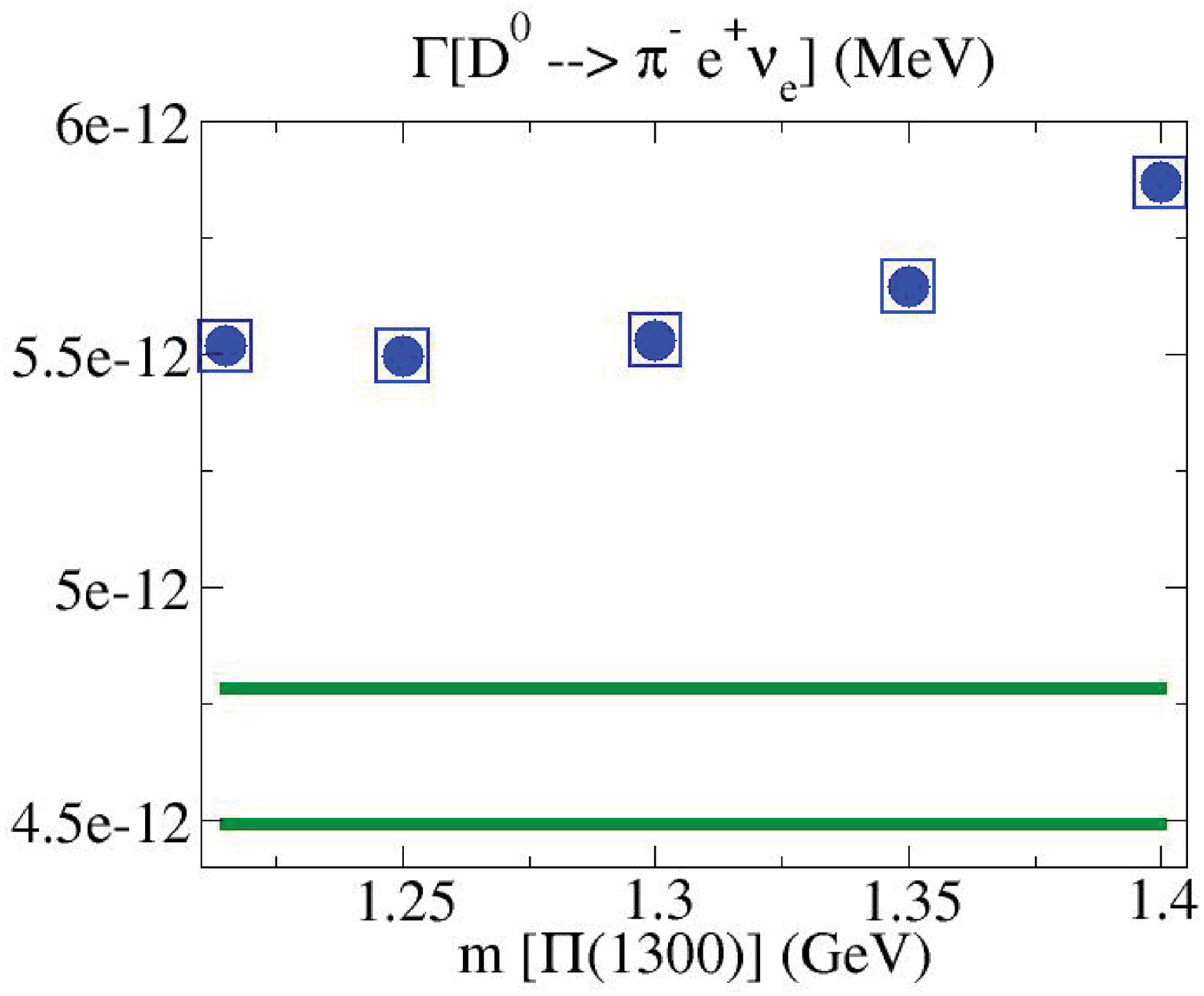}}
{\includegraphics[width=7cm,clip=true]{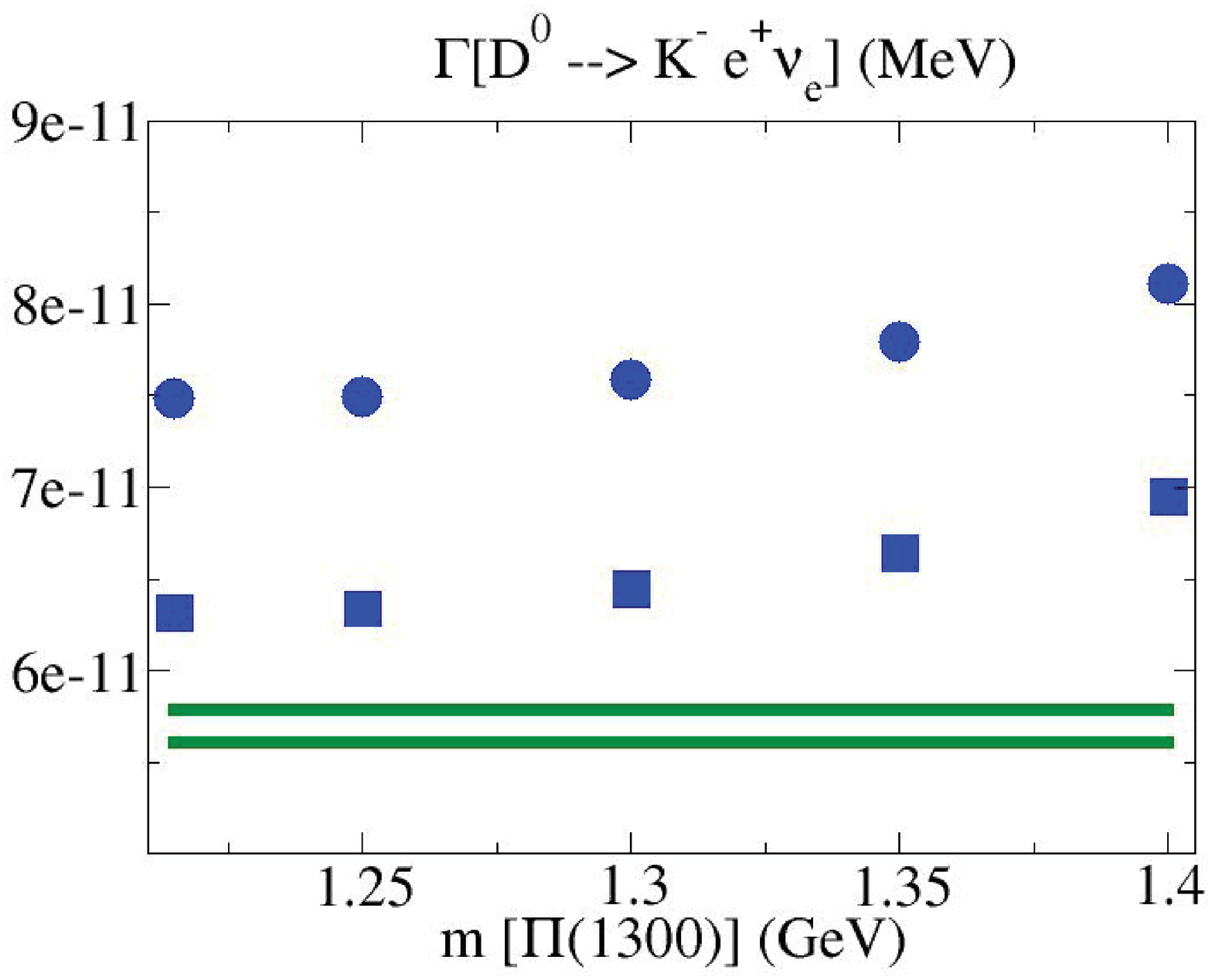}}
\caption
{Prediction of the $MM'$ model for the  partial decay width of $D^0 \rightarrow \pi^- e^+ \nu_e$ (left) and $D^0 \rightarrow K^- e^+ \nu_e$ (right) versus $m[\pi(1300)]$ for values of $A_3/A_1$ equal to 20 (circles) and 30 (squares).   The horizontal lines show the experimental range \cite{pdg}.}
\label{D0_piKev_R}
\end{figure}

The schematic diagram for this decay is shown in  Fig. \ref{D0_piKev_FD}.   The decay proceeds via the vector current described in Eq.  (\ref{heavycurrents}).     For the $\pi^-$ channel,
\begin{equation}
V_{\mu 4}^2=i\phi_4^1\stackrel{\leftrightarrow}{\partial_\mu}\phi_1^2 + \cdots =
i (R_\pi)_{11} D^0  {\partial_\mu} \pi^+ + \cdots
\end{equation}
where $R_\pi$ is the rotation matrix for the isotriplet pseudoscalars computed in \cite{global} that projects the $d {\bar u}$ pair onto the wave function of the $\pi^-$.   For the $K^-$ channel,
\begin{equation}
V_{\mu 4}^2=i\phi_4^1\stackrel{\leftrightarrow}{\partial_\mu}\phi_1^3 + \cdots =
i (R_K)_{11} D^0  {\partial_\mu} K^+ + \cdots
\end{equation}
where $R_K$ is the rotation matrix for the isodoublet pseudoscalars computed in \cite{global} that projects the $d {\bar u}$ pair onto the wave function of the $K^-$.   The model predictions for both of these two channels are presented in Fig. \ref{D0_piKev_R}, and in both cases, there is an order of magnitude agreement.

\section{Summary and discussion}
The main objective of the present work was to further test the generalized linear sigma model of ref. \cite{global} that provides a global picture for the underlying two- and four-quark components of scalar and pseudoscalar mesons below and above 1 GeV.   While a precision calculation of the partial decay widths for several semileptonic decays of $D$ mesons was not the primary objective,  we saw that the global picture presented in \cite{global} predicts these decay widths in a reasonable agreement with the available experimental data.   We interpret this qualitative agreement as further support for the global picture of scalars and pseudoscalars presented in \cite{global}.    The calculations were done in the leading order of the model.   We expect that additional corrections, such as, for example,  the effect of higher order terms in the potential, or addition of scalar and pseudoscalar gluballs, will further refine these predictions.    This framework can be applied to processes such as
$B_c^+ \rightarrow scalar + e^+ + \nu_e$ that might be useful for learning about mixing between
a $c\bar{c}$ scalar and the lighter three flavor scalars, or $B_s^0\rightarrow J/\psi f_0(980)$ measured by LHCb.

 \section*{Acknowledgments} \vskip -.5cm
  The work of
 J.Schechter was supported in part by the U.
S. DOE under Contract no. DE-FG-02-85ER 40231.  The work of R. J. was supported by a grant of the Ministry of National Education, CNCS-UEFISCDI, project number PN-II-ID-PCE-2012-4-0078.

\appendix

\section{Kinematics}
The usual weak interaction Lagrangian is,
\begin{equation}
{\cal L}=\frac{g}{2\sqrt{2}}(J_{\mu}^{-}W_{\mu}^{+}+J_{\mu}^{+}W_{\mu}^{-}),
\label{eq:lag}
\end{equation}
wherein,
\begin{eqnarray}
J_{\mu}^{-} & = & i\bar{U}\gamma_{\mu}(1+\gamma_{5})VD + i\bar{\nu}_{e}\gamma_{\mu}(1+\gamma_{5})e,\nonumber \\
J_{\mu}^{+} & = & i\bar{D}\gamma_{\mu}(1+\gamma_{5})V^{\dagger}U + i\bar{e}\gamma_{\mu}(1+\gamma_{5})\nu_e.
\label{eq:curr}
\end{eqnarray}
Here the column vectors of the quark fields take the form:
\begin{equation}
U=\left[\begin{array}{c}
u\\
c\\
t\end{array}\right],\quad\quad D=\left[\begin{array}{c}
d\\
s\\
b\end{array}\right],
\label{eq:udcol}
\end{equation}

and the CKM matrix, $V$ is explicitly,

\begin{equation}
V=\left[\begin{array}{ccc}
V_{ud} & V_{us} & V_{ub}\\
V_{cd} & V_{cs} & V_{cb}\\
V_{td} & V_{ts} & V_{tb}\end{array}\right].
\label{eq:ckm}
\end{equation}

The corresponding semi-leptonic decay amplitudes are thus,
\begin{equation}
amp\left[
          \phi_4^i (p) \rightarrow \Phi_\alpha^i(q)  +e^+(k)+\nu_e(l))
    \right]
=-i\frac{G_{F}}{\sqrt{2}}V_{c\alpha} R \left( \phi_\alpha^i \rightarrow \Phi_\alpha^i \right)
\langle \phi_\alpha^i(q)|V_{\mu4}^{\alpha}|\phi_4^i(p)\rangle
\times \bar{u}({\bf l}) \gamma_{\mu}(1+\gamma_{5})v(\bf{k}),
\label{eq:amp}
\end{equation}
where $\alpha=2,3$ (2 for $d$ quark and 3 for $s$ quark); $i$=1..3 ($u$, $d$ and $s$). This means $\phi_4^1=D^0$, $\phi_4^2=D^+$ and $\phi_4^3=D_s^+$.    The physical pseudoscalar states $\Phi_i^\alpha$ are related by appropriate rotation matrices $R\left( \phi_\alpha^i \rightarrow \Phi_\alpha^i \right)$ to the two- and four-quark components in nonets $\phi$ and $\phi'$.   For example,  $\Phi_2^1=\pi^-$,  $\phi_2^1 \propto d {\bar u}$ and $\Phi_2^1 = \left(R_\pi\right)_{11} \phi_2^1$, etc.   The spinor $v(\bf{k})$ represents the outgoing $e^+$ and $\bar{u}(\bf{l})$ represents the outgoing $\nu_e$.

The squared amplitudes, summed over the emitted lepton's spins, are then,
\begin{equation}
G_F^2|V_{c\alpha}|^2 \frac{1}{m_e^2}
\left[ R\left( \phi_\alpha^i \rightarrow \Phi_\alpha^i \right)\right]^2
[2k\cdot(p+q){l\cdot(p+q)}-l\cdot k(p+q)^2],
\label{sqamp}
\end{equation}
wherein $m_e$ has been set to zero except for the overall $1/m_e^2$ factor.  This yields the unintegrated decay width,
\begin{equation}
\frac{d\Gamma}{d|\bf{q}|}= \frac{G_F^2|V_{c\alpha}|^2}{12\pi^3}
\left[ R\left( \phi_\alpha^i \rightarrow \Phi_\alpha^i \right)\right]^2
 m(\phi_4^i)\frac{|{\bf q}|^4}{q_0}.
\label{udw}
\end{equation}

For integrating this expression we need,
\begin{equation}
|q_{max}|=\frac{m^2(\phi_4^i)-m^2(\Phi_\alpha^i)}{2m(\phi_4^i)},
\label{qmax}
\end{equation}
The indefinite integral formula, where $x=|\bf{q}|$,
\begin{equation}
\int \frac{x^4dx}{\sqrt{x^2+m^2(\Phi_\alpha^i)}}=\frac{x^3}{4}\sqrt{x^2+m^2(\Phi_\alpha^i)}-\frac{3}{8}
m^2(\Phi_\alpha^i) x\sqrt{x^2+m^2(\Phi_\alpha^i)}+
\frac{3}{8}m^4(\Phi_\alpha^i) {\rm ln} (x+\sqrt{x^2+m^2(\Phi_\alpha^i)}).
\label{integral}
\end{equation}

\end{document}